# Nonlocal Interferometry Using Macroscopic Coherent States and Weak Nonlinearities

B.T. Kirby and J.D. Franson

*Physics Department, University of Maryland, Baltimore County, Baltimore, MD 21250*

A method for performing nonlocal interferometry using phase-entangled macroscopic coherent states is described. The required entanglement can be generated using weak nonlinearities while Bell's inequality can be violated using single photons as a probe. The entanglement is relatively robust against photon loss and Bell's inequality can be violated over a relatively large distance in optical fibers despite the fact that a large number of photons are absorbed in the process.

## I. Introduction

Nonlocal interferometers can violate Bell's inequality, which is of fundamental interest in addition to being of practical use in quantum communications. Here we describe a macroscopic generalization of the nonlocal interferometer previously introduced by one of the authors [1]. In this approach, weak nonlinearities [2,3] are used to generate phase entanglement between macroscopic coherent states while single photons are used to probe the entanglement in such a way as to violate Bell's inequality. Large numbers of photons can be absorbed from these macroscopic entangled states with only a relatively small reduction in the visibility, which should allow violations of Bell's inequality over relatively large distances.

There has been considerable interest recently in methods for producing macroscopic entangled states using single-photon displacement operations [4-6] or phase-covariant cloning [7,8]. The degree of entanglement in those systems can be measured in various ways, such as by using an entanglement witness. The approach described here allows the nonlocal nature of the macroscopic entanglement to be observed as a violation of Bell's inequality.

Under ideal conditions, the nonlocal interferometer described here can produce a maximum violation of the Clauser-Horne-Shimony-Holt (CHSH) form of Bell's inequality as described in Section II. Photon loss will produce some amount of decoherence as a result of entanglement between the macroscopic states and the environment. The loss in visibility due to decoherence of that kind is analyzed in Section III, where it is found that large numbers of photons can be absorbed by the environment with a relatively small decrease in the nonlocal interference. Photon loss will also increase the overlap of two coherent states with different phase shifts, which can further reduce the observed visibility as discussed in Section IV. These effects are combined in Section V to calculate the expected visibility of the interference pattern as a function of the distance of propagation in optical fibers. A summary and conclusions are provided in Section VI.

## II. Nonlocal interferometer

Phase entanglement can be produced between two macroscopic coherent states using the single-photon interferometer illustrated in Fig. 1. Cross-phase modulation from nonlinear (Kerr) media located in each path through the interferometer will produce a small nonlinear phase shift of $2\phi$ in one of two laser beams depending on the path taken by the single photon [9-11]. A constant phase shift is applied to both beams so that the coherent states will undergo a net phase shift of $\pm\phi$. Nemoto and Munro [2,3] showed that a single photon can produce a nonlinear phase shift that is sufficiently large for the phase-shifted coherent states to be nearly orthogonal if their amplitudes are large, as illustrated in Fig. 2. A nonlinear phase shift could be produced using the giant Kerr effect [12-14] from electromagnetically-induced transparency, for example.

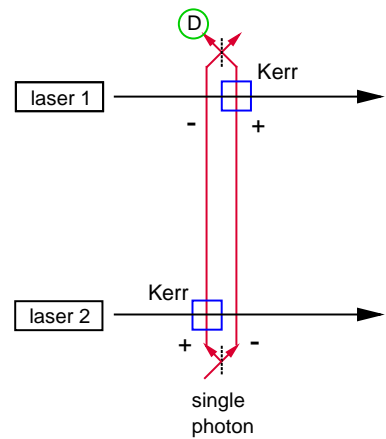

Fig. 1. (Color) Two macroscopic coherent states (laser beams) pass through a Kerr medium and experience a nonlinear phase shift of $2\phi$ when a single photon also passes through the corresponding medium. A constant phase shift is applied to both beams so that the coherent states will undergo a net phase shift of $\pm\phi$. The dashed lines represent beam splitters and the output of the interferometer is post-selected for those cases in which the single-photon detector D registers a count. This produces a coherent superposition of phase-entangled states as illustrated in Fig. 2.

The two coherent states at the output of the device shown in Fig. 1 are only accepted if the single photon triggers detector D, as indicated by the circle in the figure. It can be seen that the phase shifts in the two coherent states are anti-correlated, as illustrated in Fig. 2. This procedure creates a coherent superposition of these two phase-shifted states, which corresponds to an entangled Schrodinger cat state [9,10,15]. The ability of a Kerr medium with a third-order nonlinear susceptibility $\chi^{(3)}$ to create phase shifts of this kind is discussed in Ref. [3].

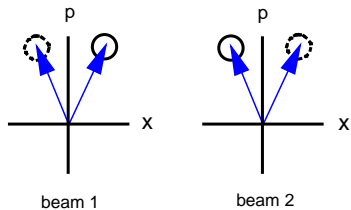

Fig. 2. (Color) Phase-space diagram of the two coherent states of Fig. 1 after their interaction with the nonlinear Kerr media. The coordinates $x$ and $p$ are the arguments of the Wigner distribution $W(x,p)$ and they correspond to the real and imaginary parts of the electric field in this case. The system is left in a superposition of macroscopic states in which beam 1 has undergone a positive phase shift while beam 2 has undergone a negative phase shift (solid circles), or vice-versa (dashed circles).

The phase-entangled source of Fig. 1 can be combined with two distant interferometers to form a nonlocal interferometer as shown in Fig. 3. A second nonlinear phase shift is applied to each beam depending on the path taken by photons B and C. A constant phase shift is also applied once again so that beams 1 and 2 will undergo a shift of $\pm\phi$ in all of the interferometers. In addition, linear phase shifts $\sigma_1$ and $\sigma_2$ are applied to the single photons in the paths indicated in Fig. 3.

The phases of the two coherent states are measured using homodyne techniques after they interact with the single photons. We will be interested in the probability $P$ of a coincidence event in which both homodyne measurements indicate zero net phase shift and the single photons trigger detectors 1, 3 and 5, as indicated by the circles in Fig. 3.

The initial coherent state of laser 1 will be denoted by $|\alpha\rangle$ while that of laser 2 will be denoted by $|\beta\rangle$. The output states of the single-photon interferometers will be denoted by $|0\rangle_i$ or $|1\rangle_i$ depending on the number of photons in path $i$. Since each coherent state is phase-shifted twice before the homodyne measurement, there are three possible final phases that the coherent states can have. The positive/negative and negative/positive phase shifts will cancel out to give zero net phase shifts, and these will be denoted by $|\alpha_{+-}\rangle = |\alpha_{-+}\rangle = |\alpha\rangle$ in beam 1.

The case of two positive phase shifts will be denoted by $|\alpha_{++}\rangle$ while two negative phase shifts will be denoted by $|\alpha_{--}\rangle$. Similar notation will be used for beam 2.

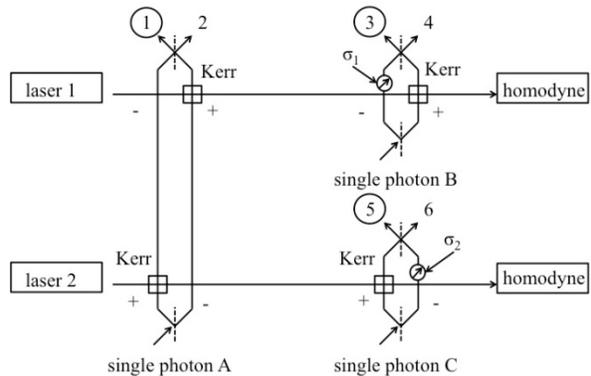

FIG. 3. Nonlocal interferometer based on phase-entangled macroscopic coherent states. A Kerr medium applies a nonlinear phase shift depending on the path taken by photons A, B, and C. Constant phase shifts are applied in each beam so that laser beams 1 and 2 undergo a net phase shift of $\pm\phi$ in each interferometer as indicated by the + and – signs. Variable (linear) phase shifts denoted by $\sigma_1$ and $\sigma_2$ are applied in interferometers B and C. We are interested in the probability of a coincidence event in which all three of the circled detectors (1, 3, and 5) are triggered and both homodyne measurements show zero net phase shifts.

After passing through all six beam splitters, the final state $|\psi\rangle$ of the system is given by

$$|\Psi\rangle = \frac{1}{2^3}[e^{i\sigma_2}|\alpha_{++}\rangle|\beta_{--}\rangle - |\alpha_{++}\rangle|\beta_{-+}\rangle$$
$$- e^{i(\sigma_1+\sigma_2)}|\alpha_{+-}\rangle|\beta_{--}\rangle + e^{i\sigma_1}|\alpha_{+-}\rangle|\beta_{-+}\rangle$$
$$- e^{i\sigma_2}|\alpha_{-+}\rangle|\beta_{+-}\rangle + |\alpha_{-+}\rangle|\beta_{++}\rangle \quad (1)$$
$$+ e^{i(\sigma_1+\sigma_2)}|\alpha_{--}\rangle|\beta_{+-}\rangle - e^{i\sigma_1}|\alpha_{--}\rangle|\beta_{++}\rangle]$$
$$\times |1\rangle_1|0\rangle_2|1\rangle_3|0\rangle_4|1\rangle_5|0\rangle_6 + |\psi_\perp\rangle.$$

This state includes a $\pi/2$ phase shift for reflections at a beam splitter. Each beam splitter doubles the number of terms in the state vector, giving a total of 64 terms in Eq. (1). We have only shown those terms in which a single photon is present in detectors 1, 3, and 5, while $|\psi_\perp\rangle$ denotes the remaining terms that are all orthogonal to the terms of interest.

The homodyne measurements are intended to distinguish between states with zero net phase shifts, such as $|\alpha_{+-}\rangle$, and states with a phase shift of $\pm 2\phi$, such as $|\alpha_{++}\rangle$. In practice, there will always be a small error in this process because two coherent states differing by a phase shift are never completely orthogonal. The overlap

between these coherent states decreases exponentially as a function of $\alpha\phi$ and it has been shown that the error in distinguishing between them is $\leq erfc[|\alpha|\sin(2\phi)/\sqrt{2}]$ when $\alpha\phi > 1$ [3]. For the time being we will assume that $\alpha\phi$ is sufficiently large that the overlap between coherent states with different phase shifts can be neglected. The more general case in which $\alpha\phi$ has been reduced as a result of photon loss will be analyzed in Section IV.

Neglecting the small overlap with states such as $|\alpha_{++}\rangle$, the homodyne measurements can be modeled as projective measurements onto those states with zero net phase shift [3]. The corresponding projection $|p\rangle$ onto the final state of interest is then given by

$$|p\rangle = \frac{1}{2^3}[e^{i\sigma_1}|\alpha_{+-}\rangle|\beta_{-+}\rangle - e^{i\sigma_2}|\alpha_{-+}\rangle|\beta_{+-}\rangle] \qquad (2)$$
$$\times |1\rangle_1 |0\rangle_2 |1\rangle_3 |0\rangle_4 |1\rangle_5 |0\rangle_6$$

while the probability $P$ of obtaining such an event is given by

$$P = \langle p | p \rangle = \frac{1}{2^6}|e^{i\sigma_1} - e^{i\sigma_2}|^2. \qquad (3)$$

Here we have made use of the fact that $|\alpha_{+-}\rangle = |\alpha_{-+}\rangle = |\alpha\rangle$ and $|\beta_{+-}\rangle = |\beta_{-+}\rangle = |\beta\rangle$. The relatively small magnitude of $P$ reflects the fact that we are looking at a small subset of the possible events.

It should be noted that $P$ is defined here as the joint probability that a photon will be found in detectors 1, 3, and 5 and the homodyne measurements will yield a net phase shift of zero. If we were to instead calculate the conditional probability that the homodyne measurements indicate zero phase shifts given that detectors 1, 3, and 5 have fired, then Equation (2) would become a post-selected state with a different normalization as is discussed in the Appendix.

Eq. (3) can be reduced to

$$P = \frac{1}{16}\left[\sin^2\left(\frac{\sigma_1 - \sigma_2}{2}\right)\right]. \qquad (4)$$

The nonlocal dependence on the difference of the two phase shifts leads to a predicted interference visibility of 100%, which violates the Bell inequality limit of 70.7% [16,17] as illustrated in Fig. 4. These results correspond to the case in which the interferometers are located near the source and there is negligible loss. The effects of loss and decoherence are included in the following sections.

These results have a simple interpretation if we realize that the only way that zero net phase shifts can occur in both beams is if photons B and C both traveled through the right path of their respective interferometers or if both photons traveled through the left path. This is analogous to the long-long and short-short path interference that occurs in the nonlocal interferometer previously proposed by one of the authors [1]. Rice et al. [18] proposed a different form of nonlocal interference for coherent states that was based on the use of a self-phase nonlinearity to produce an approximate state-dependent phase shift.

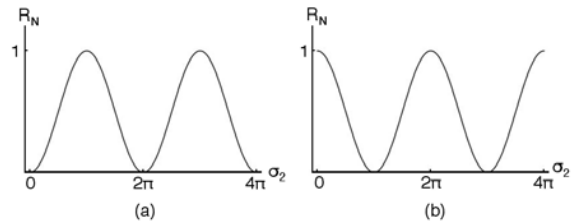

Fig. 4. Nonlocal interference pattern in the absence of loss or decoherence. (a) A plot of the normalized coincidence counting rate $R_N$ as a function of $\sigma_2$ with $\sigma_1 = 0$. (b) A plot of $R_N$ as a function of $\sigma_2$ with $\sigma_1 = \pi$. (Dimensionless units.)

### III. Photon loss and decoherence

The decoherence due to photon loss can be analyzed in several different ways. It is commonly assumed that all loss mechanisms are equivalent to inserting beam splitters in the optical paths and we will first analyze the effects of decoherence based on that assumption. We will also analyze the situation in which $N_A$ resonant two-level atoms are located in both paths between the source and the interferometers. It will be found that the decoherence due to atomic absorption of that kind is equivalent to beam splitter loss in this interferometer arrangement. Our earlier work on entangled photon holes [19] showed that beam splitter loss is not in general equivalent to absorption by resonant atoms, which illustrates the need to evaluate the effects of decoherence both ways.

First consider a single beam splitter with a small reflectivity coefficient $r$ located in the path to interferometer B, as illustrated in Fig. 5. The coherent states $|\alpha_+\rangle$ and $|\alpha_-\rangle$ will produce slightly different coherent states in the output port of the beam splitter given by

$$|\gamma_+\rangle = |r\alpha_0 e^{i\phi}\rangle$$
$$|\gamma_-\rangle = |r\alpha_0 e^{-i\phi}\rangle \qquad (5)$$

where $\alpha_0$ is the amplitude of the original coherent state. This causes the coherent states that propagate towards the interferometer to become entangled with the field in the output port of the beam splitter, so that in the overall state of the system those terms are replaced by

$$|\alpha_+\rangle \to |\alpha'_+\rangle|\gamma_+\rangle$$
$$|\alpha_-\rangle \to |\alpha'_-\rangle|\gamma_-\rangle. \qquad (6)$$

Here the primes denote the fact that the beam splitter loss will also reduce the amplitude of the coherent states as discussed in more detail in the next section.

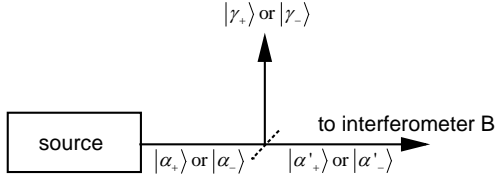

Fig. 5. Decoherence due to a beam splitter (dashed line) with a small reflectivity coefficient $r$ inserted into the path to interferometer B. A coherent state $|\alpha_+\rangle$ with a small positive phase shift will produce a weak coherent state $|\gamma_+\rangle$ in the output port of the beam splitter with the corresponding phase, while a coherent state $|\alpha_-\rangle$ with a negative phase will produce an output state $|\gamma_-\rangle$. The fact that the coherent states arriving at interferometer B are entangled to slightly different beam-splitter states produces a small amount of decoherence.

If we assume that a similar beam splitter is also included in the path to interferometer C, then Eq. (2) becomes

$$|p\rangle = \frac{1}{2^3}[e^{i\sigma_1}|\alpha'_{+-}\rangle|\beta'_{-+}\rangle|\gamma_+\rangle|\delta_-\rangle$$
$$-e^{i\sigma_2}|\alpha'_{-+}\rangle|\beta'_{+-}\rangle|\gamma_-\rangle|\delta_+\rangle]|1\rangle_1|0\rangle_2|1\rangle_3|0\rangle_4|1\rangle_5|0\rangle_6. \qquad (7)$$

Here $|\delta_+\rangle$ and $|\delta_-\rangle$ denote the corresponding coherent states in the output port of the beam splitter in the path to interferometer C. When we evaluate $P = \langle p | p \rangle$, the cross-terms that are responsible for the interference effects will be reduced by a factor of $f^2$ where $f$ is given by

$$f = \langle \gamma_-|\gamma_+\rangle = \langle \delta_-|\delta_+\rangle. \qquad (8)$$

The magnitude of the inner product of the two coherent states in the output port of one of the beam splitters is given by [20]

$$|f|^2 = |\langle \gamma_+|\gamma_-\rangle|^2 = \exp[-|\gamma_+ - \gamma_-|^2], \qquad (9)$$

where $\gamma_+ - \gamma_- = r\alpha_0(e^{i\phi} - e^{-i\phi})$ from Eq. (1). If $\phi \ll 1$ then a Taylor series expansion of the $\exp[\pm i\varphi]$ terms can be inserted into Eq. (9) to give

$$|f| = \exp[-2(r\alpha_0\phi)^2] = \exp[-2N_L\phi^2] \qquad (10)$$

Here $N_L = (r\alpha_0)^2$ is the mean number of photons lost in each beam. The imaginary part of $f$ corresponds to a small phase shift that can be compensated experimentally if necessary by adjusting the phases of the relevant interferometers. This has no effect on the visibility of the nonlocal interference and $|f|$ determines the effects of decoherence. If multiple beam splitters are included in the two paths, then Eq. (10) is still valid with $N_L$ equal to the total loss in each path due to the multiplicative properties of exponentials.

The decoherence due to atomic absorption can be analyzed in a similar way if we assume that $N_A$ resonant two-level atoms are located in the path between interferometers B and C instead of beam splitters. The interaction between the atoms and a beam of light whose phase has been shifted by $\phi$ will produce a small probability amplitude for a transition between the ground state $|G_j\rangle$ and excited state $|E_j\rangle$ of atom $j$ as described by

$$|G_j\rangle \to (1-\varepsilon^2/2)|G_j\rangle + i\varepsilon e^{i\phi}|E_j\rangle. \qquad (11)$$

Here $\varepsilon$ is a small parameter related to the atomic matrix elements in a perturbative treatment. With the inclusion of the atoms, the first term on the right-hand side of Eq. (2) should include an outer product with an atomic state $|A\rangle$ given by

$$|A\rangle = \prod_{j=1}^{N_A}\left[(1-\varepsilon^2/2)|G_j\rangle + i\varepsilon e^{i\phi}|E_j\rangle\right]. \qquad (12)$$

The second term in Eq. (2) should include the same factor with the sign of $\phi$ reversed.

This entanglement between the field and the state of the atoms contains which-path information that will reduce the visibility of the nonlocal interference. Including the factors of $|A\rangle$ in Eqs. (2) and (3) will reduce the magnitude of the cross-terms responsible for the quantum interference by a factor $f$ that is now given by

$$|f| = \prod_{j,k=1}^{N_A}\text{Re}\big[\big((1-\varepsilon^2/2)\langle G_k| - i\varepsilon e^{i\phi}\langle E_k|\big) \\ \times \big((1-\varepsilon^2/2)|G_j\rangle + i\varepsilon e^{i\phi}|E_j\rangle\big)\big]. \qquad (13)$$

In the limit of large $N_A$ and small $\varepsilon$ and $\phi$ this reduces to

$$|f| = \exp[-2N_A \varepsilon^2 \phi^2] = \exp[-2N_L \phi^2]. \qquad (14)$$

where $N_L$ is the total number of photons absorbed on average in each of the beams. This result can be seen to be equivalent to the effects of beam splitter loss when expressed in terms of the total number of photons lost in the two beams.

The visibility v of an interference pattern is defined as

$$v = \frac{R_{max} - R_{min}}{R_{max} + R_{min}} \qquad (15)$$

where $R_{max}$ and $R_{min}$ are the maximum and minimum counting rates. By inserting the appropriate factors of $f$ into the cross-terms in Eq. (3) it is straightforward to show that the visibility is reduced to $v = |f|^2$.

It can be seen that a large number of photons can be absorbed with only a modest loss in visibility if the magnitude of $\phi$ is sufficiently small. Additional insight can be obtained by writing $N_L$ in the form

$$N_L = gN_0 = g\alpha_0^2. \qquad (16)$$

Here $N_0$ is the initial number of photons and $g$ is the fraction of photons that are lost as a result of beam splitters or atomic absorption. Eqs. (10) and (14) then become

$$|f| = \exp[-2g(\alpha_0\phi)^2]. \qquad (17)$$

The fractional loss $g$ has a maximum value of 1 and it can be seen that the decoherence factor approaches a constant value in the limit of large propagation distances rather than dropping off exponentially to zero.

A nonlinear phase shift of 0.28 radians has been demonstrated at the single-photon level using a resonant cavity and a single atom [21], and comparable results have been predicted using an atomic vapor [12-14]. If we conservatively assume that $\phi = 0.014$ radians and take $\alpha = 100$ for example, then a total loss of 500 photons would reduce the visibility to 82%. This is still sufficient to violate Bell's inequality and would allow this approach to be used to implement quantum communications protocols of several kinds.

The effects of larger photon loss are illustrated in Fig. 6, which corresponds to a total loss of 4000 photons. These results neglect the fact that the overlap of coherent states with different phase shifts will increase as their amplitude decreases, as is taken into account in the next section. Thus Fig. 6 illustrates the effects of decoherence alone. The effects of decoherence on superpositions of coherent states have also been studied by Jeong [22] within the context of quantum computing.

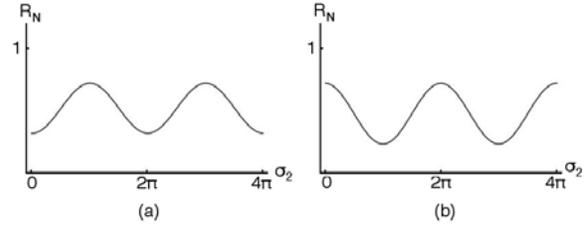

Fig. 6. Nonlocal interference pattern including the decoherence due to a total loss of 4000 photons. (a) A plot of the normalized coincidence counting rate $R_N$ as a function of $\sigma_2$ with $\sigma_1 = 0$. (b) A plot of $R_N$ as a function of $\sigma_2$ with $\sigma_1 = \pi$. These results correspond to $\phi = 0.01$ radians and an initial value of $\alpha = 100$. The effects of the reduction in the magnitude of $\alpha$ have not been included here and these results illustrate the effects of decoherence only. (Dimensionless units.)

**IV. Overlap of coherent states**

It is well known that a coherent state that undergoes linear loss will remain a coherent state but with a reduced amplitude. Consider a superposition of coherent states with different phases that are nearly orthogonal initially as illustrated in Fig. 7. As the coherent states undergo loss they will move towards the origin and begin to overlap each other. The inner product of two such coherent states will become appreciable and it is no longer valid to ignore terms such as $\langle \alpha_{+-} | \alpha_{--} \rangle$ in Eqs. (2) and (3). It is interesting to note that the process illustrated in Fig. 7 would violate unitarity if it were not for the fact that the coherent states are also becoming entangled with the environment; the factors of $f$ maintain the inner product of any two terms in the state vector.

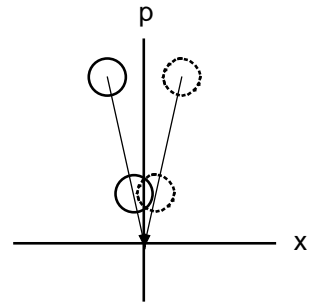

Fig. 7. Reduction in the amplitude of two coherent states with different phases. The two states are nearly orthogonal initially but they begin to overlap as their amplitudes are reduced by a linear loss mechanism such as a series of beam splitters.

A more detailed treatment of the homodyne measurements is required when coherent states with different phases begin to overlap. In particular, the probability to obtain a specific result from a homodyne measurement will now show quantum interference between the contributions from the overlapping coherent states.

We first consider the usual description of a balanced homodyne measurement on a single coherent state with no entanglement or decoherence. The local oscillator used in the homodyne measurement will define a single mode of the incident electromagnetic field that is to be measured. A single mode of the second-quantized field is mathematically equivalent to an harmonic oscillator, where the displacement and momentum operators $\hat{q}$ and $\hat{p}$ correspond to the usual linear combinations of the annihilation and creation operators $\hat{a}$ and $\hat{a}^\dagger$. Physically, $\hat{q}$ and $\hat{p}$ correspond to two phase quadratures of the field and the phase of the local oscillator can be chosen to provide a direct measurement of $\hat{q}$ for example.

The probability distribution for the measured value of $\hat{q}$ can be obtained from the usual probability amplitude $\psi(q)$ in the coordinate representation given by [20]

$$\psi(q) = \langle q|\alpha\rangle = e^{|\alpha|^2/2}\sum_{n=0}^{\infty}\frac{\alpha^n}{\sqrt{n!}}\langle q|n\rangle$$
$$= \left(\frac{\omega}{\pi\hbar}\right)^{1/4} e^{|\alpha|^2/2} e^{-\omega q^2/2\hbar} \sum_{n=0}^{\infty}\frac{\alpha^n}{n!}\frac{1}{2^{n/2}}H_n\left[\left(\frac{\omega}{\hbar}\right)^{1/2}q\right]. \quad (18)$$

Here $H_n$ is a Hermite polynomial of degree $n$, the coordinate q corresponds to the result of the homodyne measurement for an appropriate choice of local oscillator phase, and $\omega$ is the frequency of the field.

Equation (18) can be simplified using the generating function $H_k(z)$ for Hermite polynomials defined by [23]

$$\sum_{k=0}^{\infty}\frac{H_k(z)v^k}{k!} = e^{2zv-v^2}. \quad (19)$$

This allows Eq. (18) to be rewritten as

$$\psi(q) = \left(\frac{\omega}{\pi\hbar}\right)^{1/4}\exp\left\{\frac{-\omega}{2\hbar}q^2 + \left(\frac{2\omega}{\hbar}\right)^{1/2}\alpha q - \frac{1}{2}|\alpha|^2 - \frac{1}{2}\alpha^2\right\}. \quad (20)$$

In dimensionless units with $x = \sqrt{\omega/\hbar}q$ this becomes

$$\psi_\alpha(x) = \left(\frac{1}{\pi}\right)^{1/4}\exp\left\{-\frac{x^2}{2} + \frac{2x\alpha}{\sqrt{2}} - \frac{1}{2}|\alpha|^2 - \frac{1}{2}\alpha^2\right\}. \quad (21)$$

Here a subscript $\alpha$ has been added to indicate the dependence of the function on $\alpha$.

It can be seen that the probability amplitude $\psi_\alpha(x)$ is a Gaussian as is well known. The function $\psi_\alpha(x)$ is often used to calculate the Wigner distribution, but it will be more convenient here to use $\psi_\alpha(x)$ itself, as was done in Ref. [15] for example. The probability density $\rho(x)$ for obtaining the value $x$ from a single homodyne measurement is then given [10] by

$$\rho(x) = \psi^*(x)\psi(x). \quad (22)$$

We will now generalize this approach to consider the phase entangled coherent states described above with the initial parameters chosen so that $|\alpha\phi|<1$, which will give some degree of overlap between coherent states with different phases. In order to illustrate the effects of nonorthogonal coherent states, it will be assumed initially that no loss or decoherence has occurred. Under these conditions we can generalize Eqs. (18) through (22) by considering the combined probability amplitude $\psi(x_1,x_2)$ defined by

$$\psi(x_1,x_2) = \langle x_1,x_2;1,3,5|\Psi\rangle. \quad (23)$$

Here $x_1$ and $x_2$ represent the values of the homodyne measurement in beams 1 and 2. The state $|x_1,x_2;1,3,5\rangle$ corresponds to using the coordinate representation for the modes of the coherent states and the number basis for the single photons in detectors 1, 3, and 5. The joint probability density $\rho(x_1,x_2)$ for the corresponding results of the homodyne measurements is then given by

$$\rho(x_1,x_2) = \psi^*(x_1,x_2)\psi(x_1,x_2). \quad (24)$$

The first eight terms in Eq. (1) must all be retained now with the result that

$$\psi(x_1,x_2) = \frac{1}{2^3}\big[e^{i\sigma_2}\psi_{\alpha++}(x_1)\psi_{\beta--}(x_2) - \psi_{\alpha++}(x_1)\psi_{\beta-+}(x_2)$$
$$- e^{i(\sigma_1+\sigma_2)}\psi_{\alpha+-}(x_1)\psi_{\beta--}(x_2) + e^{i\sigma_1}\psi_{\alpha+-}(x_1)\psi_{\beta-+}(x_2)$$
$$- e^{i\sigma_2}\psi_{\alpha-+}(x_1)\psi_{\beta+-}(x_2) + \psi_{\alpha-+}(x_1)\psi_{\beta++}(x_2)$$
$$+ e^{i(\sigma_1+\sigma_2)}\psi_{\alpha--}(x_1)\psi_{\beta+-}(x_2) - e^{i\sigma_1}\psi_{\alpha--}(x_1)\psi_{\beta++}(x_2)\big]. \quad (25)$$

Subscripts such as $\alpha++$ denote a coherent state that has undergone two positive phase shifts, for example, and Eq. (21) is to be evaluated accordingly. Once again, $\rho(x_1, x_2)$ corresponds to a joint probability distribution and the normalization of Eq. (25) would be different for a post-selected state as discussed in the Appendix.

The expression for $\rho(x_1, x_2)$ contains a total of 64 terms and was evaluated numerically. Fig. 8 shows a plot of the probability distribution of $x_1$ and $x_2$ for the case where $\alpha\phi = 2.0$, which is sufficiently large that there is negligible overlap between the coherent states. Various peaks corresponding to coherent states whose phases were shifted by $\pm 2\phi$ can be seen. The interferometer protocol selects only those events in the vicinity of $x_1 = x_2 = 0$ which corresponds to zero net phase shift. Fig. 8a corresponds to interferometer phase settings with $\sigma_1 - \sigma_2 = 0$, which is expected to minimize the probability of obtaining values of $x_1 = x_2 = 0$. It can be seen that the probability of such an event is nearly zero when there is minimal overlap between the coherent states. Fig. 8b shows the corresponding probability distribution for interferometer settings of $\sigma_1 - \sigma_2 = \pi$ which maximizes the probability of obtaining a result near $x_1 = x_2 = 0$ as expected. Once again, these results do not include any decoherence due to photon loss.

In Fig. 9, the value of $\alpha\phi$ was reduced to 0.3, which gives a significant overlap between the different coherent states. The peaks in the probability distribution all move closer to the origin as would be expected from Fig. 7. In addition, quantum interference between the superposition of coherent states is evident near the origin. The overlap of the coherent states reduces the amount of true nonlocal interference and it adds some amount of local single-photon interference that tends to mask the nonlocal effects of interest.

The effects of decoherence due to photon loss can be included in this analysis by adding entanglement with the states at the output port of a beam splitter as in Eq. (7) or atomic states as in Eq. (12). This has the effect of adding factors of $f^2$ to some of the 64 terms in $\rho(x_1, x_2)$, which was also evaluated numerically. Fig. 10 shows a plot of the probability distribution for $\sigma_1 - \sigma_2 = 0$, which would produce zero probability of obtaining homodyne measurements with $x_1 = x_2 = 0$ in the ideal case. It can be seen that photon loss decreases the amplitude of the coherent states and will eventually cause them to overlap. In addition, the decoherence due to entanglement with the environment reduces the visibility of the nonlocal interference effects. As the decoherence increases, the magnitude of the central peak increases from its ideal value of zero.

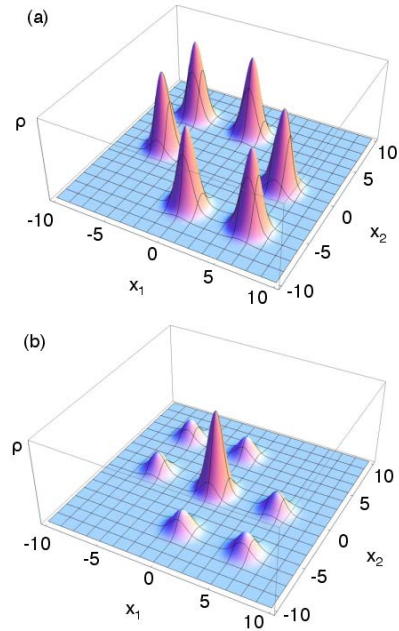

Fig. 8. (Color) Plots of the probability density $\rho(x_1, x_2)$ as a function of the quadrature measurements $x_1$ and $x_2$ from balanced homodyne measurements on beams 1 and 2. These results correspond to the ideal case with negligible overlap between the coherent states and no decoherence. (a) Setting the interferometer phases to give $\sigma_1 - \sigma_2 = 0$ produces nearly zero probability of obtaining $x_1 = x_2 = 0$. (b) Setting the interferometer phases to give $\sigma_1 - \sigma_2 = \pi$ produces the maximum probability of obtaining $x_1 = x_2 = 0$. These results correspond to $\alpha = 100$ and $\phi = 0.02$. (Dimensionless units.)

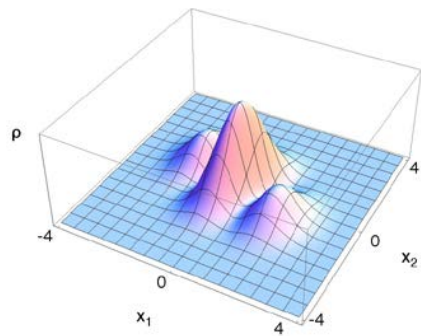

Fig. 9. (Color) A plot of the probability density $\rho(x_1, x_2)$ as in Fig. 8b, but with $\alpha\phi$ reduced to 0.3 to give a significant overlap between coherent states with different phase shifts. Here $\sigma_1 - \sigma_2 = \pi$ as in Fig. 8b but now the effects of quantum interference between the coherent states is apparent. These results correspond to $\alpha = 100$ and $\phi = 0.003$ and it was assumed once again that there is no decoherence. (Dimensionless units.)

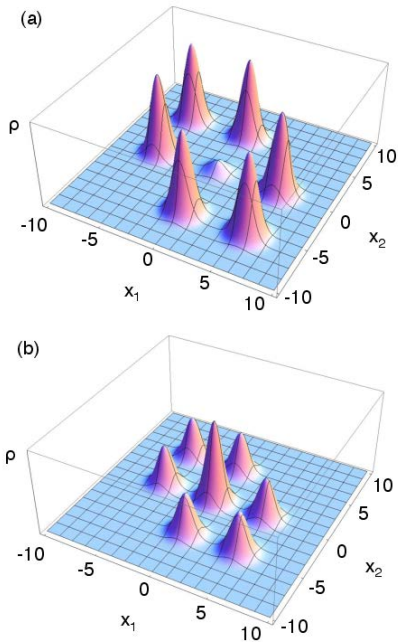

Fig. 10. (Color) A plot of the probability density $\rho(x_1, x_2)$ with $\sigma_1 - \sigma_2 = 0$ as in Fig. 8a. These results include the combined effects of decoherence and the overlap of the coherent states due to photon loss. (a) Effects of decoherence and overlap after the loss of 100 photons. (b) Effects of decoherence and overlap after the loss of 5800 photons. All of these results correspond to an initial value of $\alpha = 100$ and $\phi = 0.02$. It can be seen that photon loss moves the coherent states closer to the origin and reduces the visibility of the interference pattern. (Dimensionless units.)

These numerical calculations can be used to determine the accuracy of the approximation made in Section II that the overlap of the coherent states can be neglected when $\alpha\phi > 1$ [3]. It was found that the visibility calculated using that approximation and Eq. (2) was the same as that obtained when the overlap was included (as in this section) to within 1% when $\alpha\phi \geq 1$. This justifies the approach used in Section II for $\alpha\phi \geq 1$.

**V. Range in Optical Fibers**

The ability to violate Bell's inequality with a total separation S between the two interferometers was determined by calculating the parameter s that appears in the Clauser-Horne-Shimony-Holt form of Bell's inequality [16]. A value of $|s| > 2$ cannot be achieved by any local hidden-variable theory and is an indication of nonlocal interference. As discussed in Section III, it is expected that a value of $\phi = 0.01$ could be routinely achieved in a cavity-based approach. In that case a coherent state amplitude of $\alpha \sim 100$ would be sufficient to limit the initial amount of overlap between the phase-shifted coherent states.

Fig. 11 shows a plot of s as a function of $\phi$ for several different values of the separation $D$ while $\alpha$ was held fixed at a value of 100. It can be seen that there is an optimal value of $\phi$ for any given value of $S$ and that the optimal value of $\phi$ increases with increasing range. If $\phi$ is too large then the decoherence factor $f$ will become very small at large distances. On the other hand, if $\phi$ is too small then the coherent states will overlap and the nonlocal interference will be reduced. It can be seen that Bell's inequality can be violated up to a distance of approximately 8.2 km for this example. The CHSH parameter $s$ continues to have a positive value at distances of at least 50 km although Bell's inequality is no longer violated, which suggests that some degree of correlation may persist at these longer ranges.

Based on the detection loop-hole [24], one might suspect that the use of post-selection on the single-photon and homodyne measurement results may allow a hidden-variable theory to be contrived in such a way as to reproduce the results predicted by the quantum theory [25]. This possibility can be ruled out by supposing that the homodyne measurements have been post-selected before the phase shifts $\sigma_1$ and $\sigma_2$ are chosen at random. In that case, the initial interferometer combined with the homodyne measurements can be viewed as an effective source that generates an entangled state of photons B and C [26]. Bell's inequality can then be violated with no further post-selection, which shows that no hidden-variable theory can reproduce these results.

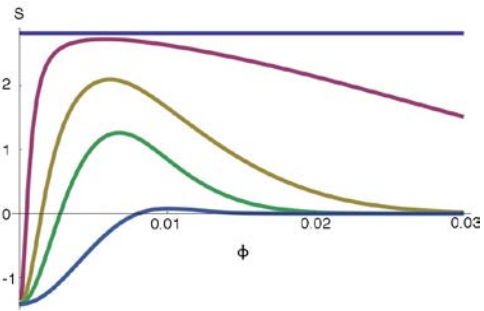

Fig. 11. (Color) A plot of the CHSH parameter $s$ as a function of the nonlinear phase shift $\varphi$ for an initial coherent state amplitude of $\alpha = 100$ and an assumed loss in optical fibers of 0.15 dB/km. The flat blue line corresponds to zero separation (no loss) while the purple line corresponds to a separation of 1 km. It can be seen that Bell's inequality can be violated for separations up to 8.2 km (yellow line). Positive values of $s$ persist out to larger separations of 20 km (green curve) and 50km (lower blue curve). (Dimensionless units.)

If the system is operated at high data rates using short pulses, then the correspondingly large bandwidth may cause dispersion in the optical fibers to become a concern. If necessary, the effects of dispersion could be minimized by limiting the bandwidth and by operating near the zero-dispersion wavelength of the optical fibers. These dispersion effects are essentially the same as in a classical communications system and methods for controlling the effects of dispersion in optical fibers already exist.

Dispersion within the nonlinear medium that is used to produce the nonlinear phase shift can pose problems that are inherently quantum mechanical, however. Shapiro and Razavi [27] have used the Kramers-Kronig relation to show that the production of a single-photon nonlinear phase shift of $\pi$ must also produce a significant amount of loss and phase noise. which may limit the use of techniques of this kind for quantum computing applications. Effects of that kind are not expected to be significant in our approach because small nonlinear phase shifts can be produced with correspondingly small amounts of loss. In addition, our approach can tolerate a significant amount of loss and still violate Bell's inequality, which is less stringent than the error threshold for quantum computing. Finally, the use of a resonant cavity with a mode spacing that is larger than the spectral bandwidth of the medium would eliminate the possibility of decay into spurious modes, which is the physical mechanism responsible for the issues raised by Shapiro and Razavi [27].

## VI. Summary and conclusions

In summary, a technique for performing nonlocal interferometry using macroscopic coherent states has been described. A single-photon interferometer combined with a weak nonlinearity (the Kerr effect) can be used to create a macroscopic phase-entangled state [9,10,15]. Two single-photon interferometers at two distant locations can then be used as a probe of the entanglement in such a way that Bell's inequality is violated. The latter procedure also makes use of a weak Kerr nonlinearity, which is used for both the generation and detection of the entanglement [28].

Macroscopic entangled states of this kind can undergo the loss of many photons and still violate Bell's inequality. Nevertheless, photon loss will reduce the amplitude of the coherent states and eventually cause them to have a significant amount of overlap. In addition, photon loss entangles the coherent states with the environment, which produces decoherence and a loss of visibility. Despite these effects, we estimate that macroscopic phase entangled states can violate Bell's inequality over relatively large distances in commercially-available optical fiber.

In addition to being of fundamental interest, these results may have applications in quantum key distribution and other forms of quantum communications. Those possibilities remain an area for further investigation.

## Acknowledgements

We would like to acknowledge valuable discussions with C. Broadbent, J. Howell, G. Jaeger, T. Pittman, S. Sergienko, and D. Simon. This work was supported in part by DARPA DSO under grant # W31P4Q-10-1-0018.

**Appendix**

The probability $P$ that is calculated in the text corresponds to the joint probability that detectors 1, 3, and 5 will detect a photon while at the same time the two homodyne measurements register a net phase shift of zero. The normalization constant in front of Eq. (1) does not change as the system evolves and the probability $P$ is correspondingly small.

An alternative way at looking at the data is to post-select on those events in which detectors 1, 3, and 5 have registered a photon. The state vector is then projected onto the corresponding subspace of Hilbert space and renormalized. One can then calculate the conditional probability $P_C$ that the homodyne measurements will register a net phase shift of zero given that the appropriate detectors have fired. In general $P_C \neq P$ although we are obviously describing the same physical phenomenon.

If the system is post-selected in this way, the normalization constant $c_n$ in front of Eq. (1) should be replaced with

$$c_n = \frac{1}{\sqrt{8 - 2\cos(\sigma_1 - \sigma_2)}}. \tag{A1}$$

This result assumes that coherent states with different phase shifts are very nearly orthogonal. It can be seen that $c_n$ is now dependent on the phase settings of the interferometers and that it oscillates around an average value of $1/(2\sqrt{2}) = 1/\sqrt{8}$ in this case.

When there is a significant overlap between the coherent states as in Section IV, the post-selected version of Eq. (25) would require a more complicated normalization that involves the inner product of 64 cross-terms. This constant can be evaluated but it is simpler to use joint probabilities as was done throughout the text.